\documentclass[english,article, aps,fleqn, twocolumn,showpacs,dvipdfmx]{revtex4}
\usepackage{amsmath}
\usepackage{amstext}
\usepackage[dvipdfmx]{graphicx}

\usepackage{ulem}
\usepackage{color}

\begin{document}

\title{Partial synchronization of relaxation oscillators with repulsive
coupling in autocatalytic integrate-and-fire model and electrochemical
experiments}

\author{Hiroshi Kori$^1$, Istv\'{a}n Z. Kiss$^2$, Swati Jain$^3$, John
L. Hudson$^{3,}$\footnote{Deceased}}
\affiliation{
$^1$ Department of Information Sciences, Ochanomizu University, Tokyo 112-8610, Japan  \\
$^2$ Department of Chemistry, Saint Louis University, St. Louis, Missouri 63103, USA  \\
$^3$ Department of Chemical Engineering, University of Virginia, Charlottesville, Virginia 22904, USA 
}
\begin{abstract}
Experiments and supporting theoretical analysis is presented to describe
the synchronization patterns that can be observed with a population of
globally coupled electrochemical oscillators close to a homoclinic,
saddle-loop bifurcation, where the coupling is repulsive in the
electrode potential.  While attractive coupling generates phase clusters
and desynchronized states, repulsive coupling results in synchronized
oscillations. The experiments are interpreted with a phenomenological
model that captures the waveform of the oscillations (exponential
increase) followed by a refractory period. The globally coupled
autocatalytic integrate-and-fire model predicts the development of
partially synchronized states that occur through attracting heteroclinic
cycles between out-of-phase two-cluster states.  Similar behavior can be
expected in many other systems where the oscillations occur close to a
saddle-loop bifurcation, e.g., with Morris-Lecar neurons.

\end{abstract}
\maketitle

\textbf{Many oscillatory processes underlie the functioning of important biological and engineered 
systems. The waveform of the oscillation of a variable (e.g., concentration of substances) has strong impact 
on the overall behavior, e.g. on how the oscillations synchronize together. The waveform can be smooth, 
nearly sinusoidal, or relaxation type, where slow variations are followed by a quick spike. In this paper, we performed 
experiments with a chemical oscillatory system, where the waveform had strong relaxation character, and show that 
such a system, in contrast with the previously studied smooth oscillation, can produce synchronization with 
repulsive coupling among the variables. The experiments are interpreted with a simple mathematical model, 
where the relaxation character of the waveform can be tuned to generate complex synchronization patterns.}

\section{Introduction}
The widespread occurrence of different types of synchronization patterns of nonlinear 
dynamical systems calls for theoretical description using simplified,
generic models \cite{winfree01,kuramoto84,pikovsky01}. 
The development 
of such models depends on local nonlinear features of the oscillating units and the type of interactions. 
When the 
interactions are global and weak, the oscillations can often be described with phase models, 
\begin{equation}
 \frac{d\phi_{i}}{dt}=\omega_{i}+\sum_{j=1}^N \Gamma(\phi_{i}-\phi_{j}),
\end{equation}
where $\phi_{i}$ $(i=1\ldots N)$ and $\omega_{i}$ are the phase
and the natural frequency of the $i$-th oscillator, respectively,
and $N$ is the number of oscillators
\cite{kuramoto84}.
The function $\Gamma$ is referred to as the phase interaction function,
obtained as the
convolution of the phase sensitivity function $Z(\phi)$ and
the waveform of an interacting agent. The function $Z(\phi)$ is
proportional to the phase response curve, which has been measured in
many systems including chemical and biological oscillators.
The central component in such a phase model 
is the functional form of $\Gamma(\phi)$. For example, close to a Hopf bifurcation with coupling 
that occurs through an additive term of variable differences, $\Gamma$
is a sinusoidal function, possibly 
shifted with a constant term, as in the Kuramoto-Sakaguchi model
\cite{kuramoto84}.
This property results from the fact that
both $Z$ and the waveform are sinusoidal \cite{kuramoto84}.
Further away from the Hopf bifurcation, 
higher harmonics can occur in $Z$, as theoretically shown \cite{kori14} and
observed in many chemical and biological oscillators
\cite{anumonwo1991phase,wessel1995vitro,khalsa2003phase,prinz2003functional,kiss05,koinuma2017transition}.
Correspondingly, higher harmonics appears also in $H$ giving rise
to phase cluster dynamics \cite{kori14}.
Because of the relative simplicity of the mathematical structure of phase models, very often 
analytical solutions exist for synchronization patterns.

Many oscillations in nature, for example, in chemistry and neurophysiology, have more complex 
shapes. The slow, exponential decaying waveform, corresponding to the charging of the membrane 
potential in biology, motivated the development of integrate-and-fire (IF) types of models. 
In IF models after the process is complete, there is quick, often instantaneous discharge that allows 
the process to restart.
Such models, which 
typically describe the behavior close to bifurcation
(e.g. homoclinic saddle loop (SL) bifurcation or  saddle-node bifurcation of
infinity period (SNIPER)) \cite{izhikevich07},
can generate rich dynamics in networks, that include synchronization \cite{kuramoto91},
asynchronous dynamics \cite{abbott93}, 
clustering \cite{kori03}, or chimera states \cite{Olmi:2010fj}.
Systems 
close to a SNIPER or SL bifurcation, can generate a refractory period: the discharging process is not instantaneous, 
but occurs over a relatively short interval during which the system is
insensitive to external perturbations \cite{Flesselles:1998fk}.

In this paper, we design an integrate-and-fire type of model for the description of synchronization 
patterns of an electrochemical oscillator close to SL bifurcation. The experiment, performed with a repulsive 
coupling of a population of electrochemical oscillators, exhibits a synchronized state. The extent 
of the synchronization is investigated as a function of distance from the SL bifurcation. The experiments 
are interpreted with an autocatalytic integrate-and-fire (AIF) model, adjusted for the
exponentially increasing waveform for the experiments.
The AIF model is analyzed using a phase model description. The model 
analysis reveals the type of synchronized oscillations, clusters, and partially synchronized states.

\section{Experimental Results and Analysis}
We carried out experiments to explore the synchronization behavior 
with $N=64$ electrochemical relaxation oscillators close to a homoclinic bifurcation. 
The system consists of 64 metal wires; the rate of metal
dissolution (currents of the electrodes) at constant circuit potential
($U$) is measured. The currents of the electrodes become oscillatory
through a supercritical Hopf bifurcation point at $U=1.0~\textrm{V}$;
the oscillations are smooth near the Hopf bifurcation point. As the
potential is increased further, relaxation oscillations are seen
that disappear into a steady state through a homoclinic bifurcation
at about $U=1.31~\textrm{V}$ \cite{kiss05}. In experiments with the circuit potential potentials 
close to the homoclinic bifurcation point, some electrodes passivated during the experiments; in this case 
these electrodes were disconnected and the experiments were continued with a smaller number of oscillators 
as indicated in the figures. In all reported experiments $N \ge 47$.
The electrodes are coupled with a combination of resistors coupled in
series ($R_{s}$) and in parallel ($R_{p}$);
the imposed coupling strength is $K=NR_{s}/R_{p}$ \cite{kiss05}.
Negative coupling
was induced with the application of negative series resistance (built
in a PAR 273A potentiostat) \cite{kori14}.

\begin{figure}
\includegraphics[clip,width=0.9\columnwidth,keepaspectratio]{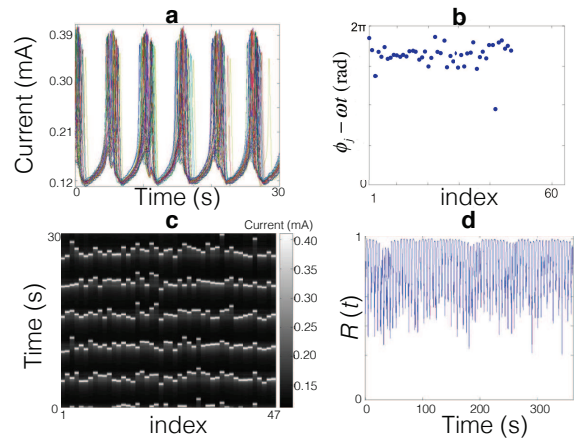}
\caption{Experiments: Nearly in-phase synchronization with negative
global coupling  very close to homoclinic bifurcation. 
(a) Times series of current oscillations. (b) Phase snapshot. (c) Grayscale plot of currents. 
(d) Order parameter vs. time.   $U=1.3$~V. }
\label{expfig1}
\end{figure}

In a previous publication \cite{kori14}, we showed that  close to the Hopf bifurcation with strong 
negative coupling the weak nonlinearities can generate cluster states. Further away from the Hopf bifurcation the system showed 
desynchronized behavior.

Here we focus on behavior at even larger circuit potentials,
where relaxation oscillations occur that 
cease through a homoclinic bifurcation.
Figure \ref{expfig1} shows that under these conditions the system exhibits 
nearly synchronized behavior. While the oscillators do not spike completely together (see Fig.~\ref{expfig1}a and c for time 
series in current vs time and grayscale plots, respectively), they form a tightly synchronized cluster. This synchronization 
can be also seen in the phase snapshot in Fig.~\ref{expfig1}b, where the phases of the oscillators, $\phi_j(t)$, were calculated with the Hilbert transform 
approach \cite{kiss02,pikovsky01}. The extent of synchronization can be
quantitatively characterized using the average Kuramoto order parameter
$\langle R \rangle$ \cite{kuramoto84}, which is obtained by averaging the order parameter 
$R(t)=\frac{1}{N}|\sum_{j=1}^N e^{{\rm i}\phi_j(t)}|$  after
a transient. We have $R(t)=1$ when all the phases take the same value
(in-phase synchrony) and $R(t)=0$ for a uniform phase distribution
including the balanced cluster states. As shown in  Fig.~\ref{expfig1}d, the synchronized oscillations generate large order parameter 
close to 1.

\begin{figure}
\includegraphics[clip,width=0.9\columnwidth,keepaspectratio]{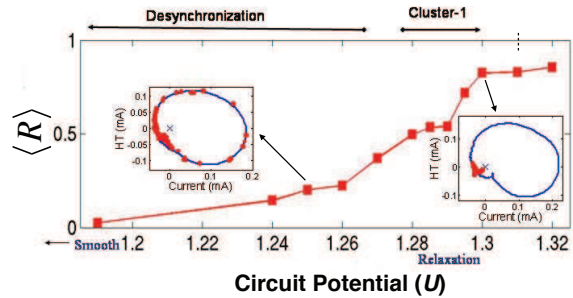}

\caption{Experiments: Emerging order with approaching saddle loop bifurcation point by increasing 
the circuit potential, $U$. The distribution of elements,
in the Hilbert transform space, at $U=1.25$~V is shown in the left
side. The distribution of elements in a synchronized
state at $U=1.3$~V is shown in right side. Dashed line indicates the position of the homoclinic bifurcation 
point of a single oscillator.}

\label{expfig2}
\end{figure}

We carried out a series of 
experiments in which the circuit potential 
was increased and the averaged Kuramoto order was calculated (see Fig.~\ref{expfig2}).
As the circuit potential (and thus the relaxation character) of the oscillators was increased, 
an increase in the Kuramoto order parameter was observed. Interestingly, the increase of the order is not very sharp, as, for example, 
could be expected from a bifurcation that leads to a stable one-cluster state. In the inset, it is shown that at $U$ = 1.25 V, 
in the state space an enhanced synchronization is present.

\subsection{Phase model analysis with experimentally obtained phase interaction function}
For better experimental characterization of the synchronization transition with increasing the circuit potential, 
we performed a phase model analysis, where the phase interaction function was
constructed from experimentally measured phase response functions \cite{kiss06,kiss05}.  Using the phase model, the stability of the different cluster states can be calculated for a population 
of globally coupled oscillators.  
For a large interval in the moderately relaxational
oscillation region, ($1.25~\textrm{V}<U<1.28~\textrm{V}$), the experimentally
measured interaction function predicted desynchrony, however, a one
cluster state with elevated value of order parameter was experimentally observed (see Fig.~\ref{expfig2}).

\begin{figure}
 \includegraphics[width=8cm]{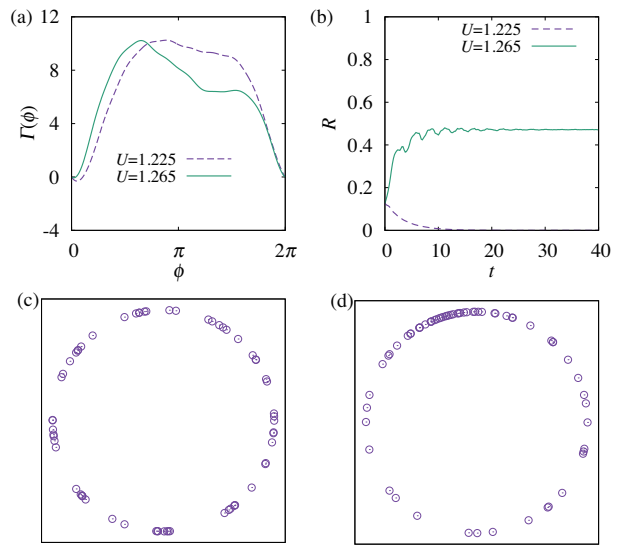}
 \caption{Numerical results using the phase model with experimentally
 measured coupling funtion. (a) phase interaction function measured experimentally for
 $U=1.225$~V and $U=1.265$~V. (b) Time-series of the order parameter
 $R(t)$. (c,d) Snapshots of the phases at $t=40$ for $U=1.225$~V and
 $U=1.265$~V, respectively. The same random initial condition was employed both
 for $U=1.225$~V and $U=1.265$~V.}
 \label{fig:exp_pm}
\end{figure}

The difference between a population generating low and elevated $\langle R \rangle$ can be demonstrated with experimentally 
determined coupling functions for $U=1.225~\textrm{V}$ and $U=1.265~\textrm{V}$ \cite{kiss06,kiss05}. The measured $\Gamma(\phi)$ functions were expressed as a
Fourier series up to tenth harmonics, which are shown in
Fig.~\ref{fig:exp_pm}(a).
By directly simulating the phase model with these $\Gamma(\phi)$ functions,
we obtained dynamics similar to experimental ones, as shown in
Fig.~\ref{fig:exp_pm}(b) and (c). 
Further using the $\Gamma(\phi)$ functions,
we checked the linear stability (see Appendix A) of the balanced $n$-cluster
states ($n=1,\ldots,10$) \cite{okuda93}, indicating that the states with $n=7,
8, 9$ are stable for $U=1.225~\textrm{V}$ whereas no state is stable for $U=1.265~\textrm{V}$.
The former result explains the almost uniform distribution of phases
observed for $U=1.225~\textrm{V}$. However, the latter can not acount for
the emergence of a type of one cluster state observed for $U=1.265~\textrm{V}$.
In order to explore the emergent collective synchronization approaching the homoclinic bifurcation, 
we develop a phenomenological model in the next section.

\section{Autocatalytic integrate-and-fire model}

\subsection{Model}
We construct a simple, one-dimensional
model for oscillations close to a homoclinic bifurcation. This model is
motivated by the waveform of many chemical and biological oscillators
composed of an 'excitatory' phase with exponential increase followed
by a usually sharp decrease with a refractory period, as observed
in Fig.~\ref{expfig1}a.
Therefore, a
coupled oscillator is described by 
\begin{equation}
 \frac{dv}{dt}=v+Kp(t).\label{model}
\end{equation}
where $v=v(t)$ is {the state variable}, $K$ is the
coupling strength, and
$p(t)$ is an external input describing the influence from
other oscillators.
Note that because the waveform is generated by a positive feedback mechanism 
in Eq.~\eqref{model}, we call the integrate-and-fire model `autocatalytic'.

\begin{figure}
 \includegraphics[width=6cm]{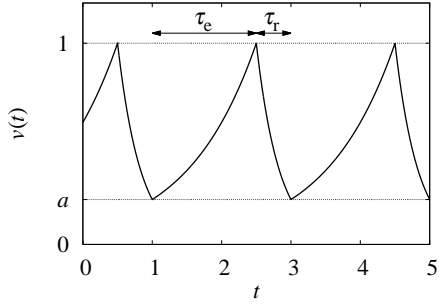}\\
 \caption{Typical waveform of the AIF oscillator. $\tau_{\rm e}=1.5,
 \tau_{\rm r}=0.5$.}
 \label{fig:aif}
\end{figure}

When $v$ reaches $1$,
its value is smoothly reset to parameter $a$ $(0<a<1)$ by obeying 
\begin{equation}
 v=e^{-b(t-t_{\textrm{fire}})},
  \label{model_reset}
\end{equation}
where $t_{\textrm{fire}}$ is the latest time at which $v$ becomes 1.
Here, we have assumed that the oscillator is not influenced by other oscillators
(or external forces) during the reseting process, i.e., in an absolute refractory period.
A typical time series in the absence of coupling (i.e. $K=0$) is
shown in Fig.~\ref{fig:aif}(a).
In this model, 
parameters $a$ and $b$ characterize the intrinsic dynamical
property of an oscillator. 
It is more convenient to characterize the relaxation
character of the oscillator by using the excitatory period $\tau_{\rm
e}$ and refractory period $\tau_{\rm r}$, given by
\begin{align}
 \tau_{\rm e} &=-\ln a,\\
 \tau_{\rm r} &=-\frac{\ln a}{b}.
\end{align}
For convenience, we denote the intrinsic period by
\begin{equation}
 T =  \tau_{\rm e} + \tau_{\rm r}.
\end{equation}
Larger $\tau_{\rm e}$ values compared to $\tau_{\rm r}$ indicate 
stronger relaxation character and
closer distance to the homoclinic bifurcation (at which $\tau_{\rm e}$
becomes infinite).

\subsection{Phase Reduction}
As a coupled oscillator system, we consider
\begin{equation}
 \frac{dv}{dt}=v+K(v'-v), \label{model2}
\end{equation}
where $v'(t)$ describes the state of an interacting oscillator.
The coupling term $K(v'-v)$ describes diffusive coupling.
In the context of electrochemical and neural dynamics,
$v$ and $K$ describes the electric potential and the conductance,
respectively. Although we have included only one interacting
oscillator described by $v'$ in Eq.~\eqref{model2} for simplicity,
we can consider a network of interacting oscillators by replacing $v'-v$
with $\sum_j (v_j - v)$, where $v_j$ are the state variables of
interacting oscillators.

The phase model corresponding to this model can be calculated 
analytically.
We first define the phase $\varphi(v(t))$ as a function of the state $v(t)$
such that $\frac{d}{dt}\varphi(v(t)) = 1$ for $K=0$,
i.e.
\begin{equation}
 \varphi(v) = \left\{
 \begin{array}{cl} 
   \ln v+\tau_{\rm e} & \quad \mbox{($0 \le \varphi < \tau_{\rm e}$, excitatory phase)},\\
   -\frac{\tau_{\rm r}}{\tau_{\rm e}} \ln v + \tau_{\rm e} & \quad \mbox{($\tau_{\rm e} \le \varphi <
    T$, refractory phase)}.          
 \end{array} \right.
\end{equation}
By solving reversely, we obtain
\begin{equation}
 v(t) = {\tilde v}(\varphi(t)),
  \label{v}
\end{equation}
where 
\begin{equation}
 {\tilde v}(\varphi) = \left\{
 \begin{array}{cl} 
  e^{\varphi-\tau_{\rm e}}  & \quad
   \mbox{($0 \le \varphi < \tau_{\rm e}$)},\\
   e^{-\frac{\tau_{\rm e}}{\tau_{\rm r}}(\varphi-\tau_{\rm e})} &
    \quad \mbox{($\tau_{\rm e} \le \varphi < T$)}.
 \end{array} \right.
\end{equation}
We now derive the dynamical equation of $\varphi(t)$.
For the excitatory phase, using $\frac{d \varphi}{dt} = \frac{d \varphi}{d
v}\frac{dv}{dt}$ with Eqs.~\eqref{model} and \eqref{v}, we obtain
\begin{align}
 \frac{d \varphi}{dt} = 1 + K \tilde Z(\varphi) \left\{ {\tilde v}(\varphi') - {\tilde v}(\varphi) \right\},
 \label{aif_pm}
\end{align}
where $\tilde Z(\varphi) = e^{\tau_{\rm e}-\varphi}$ and $\varphi'(t)$ is the
phase of the state $v'(t)$.
For the refractory phase, we have $\dot \varphi=1$.
Altogether, we obtain Eq.~\eqref{aif_pm} with $\tilde Z$ redefined as
\begin{equation}
 \label{Z}
 \tilde Z(\varphi)   = \left\{
 \begin{array}{cl} 
  e^{\tau_{\rm e}-\varphi} &
   \quad \mbox{($0 \le \varphi < \tau_{\rm e}$)},\\
   0  &  \quad \mbox{($\tau_{\rm e} \le \varphi < T$)},
 \end{array} \right.
\end{equation}
The function $\tilde Z(\varphi)$ is called the phase sensitivity,
which describes the strength and sign of response
of the phase to a perturbation applied to the oscillator.
This function is defined as $\tilde Z(\varphi) = \left(\frac{d \tilde v}{d
\varphi}\right)^{-1}$, i.e. $\left(\frac{d v}{d t}\right)^{-1}$ in
the unperturbed system. This implies that if the perturbation is given
to the oscillator when it evolves slowly, i.e., when $v$ is small in the
AIF model, the phase
response is large. Therefore, $\tilde Z(\varphi)$ decreases with increasing
$\varphi$, as described in Eq.~\eqref{Z}.

\begin{figure}
\includegraphics[width=8cm]{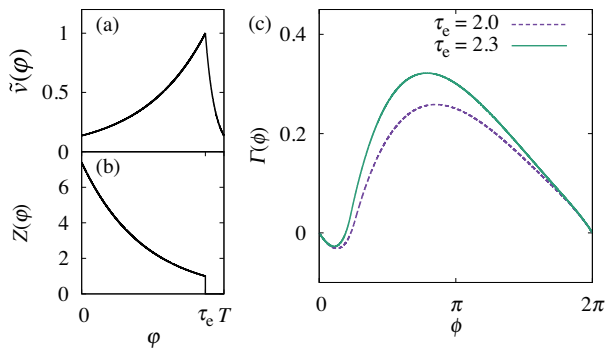}
 \caption{Functions in the AIF model. (a) Waveform ${\tilde v}(\varphi)$. (b)
 Phase sensitivity function $Z(\varphi)$. (c) Phase interaction functions
 $\Gamma(\phi)$.
 $\tau_{\rm e}=2.0$ in (a) and (b);
 and $\tau_{\rm r}=0.3$.}
 \label{fig:aif_pm}
\end{figure}

For $K \ll 1$, we may further reduce Eq.~\eqref{aif_pm} to a more
tractable equation, given as
\begin{equation}
 \dot \varphi = 1 + K \tilde \Gamma(\varphi - \varphi'),
  \label{pm1}
\end{equation}
where the coupling function $\tilde \Gamma$ is obtained by 
by averaging the right hand side
of Eq.~\eqref{aif_pm} over the period $T$ \cite{kuramoto84},
i.e.,
\begin{equation}
 \tilde \Gamma(\varphi) = \frac{1}{T} \int_{0}^T
  \tilde Z(\varphi + \theta) \left\{ {\tilde v}(\theta) - {\tilde v}(\varphi+\theta) \right\} d \theta.
\end{equation}
For $\tau_{\rm e}> \tau_{\rm r}$, which we assume henceforth, we obtain
\begin{equation}
 \tilde \Gamma(\varphi) = \left\{
 \begin{array}{cl} 
  \{\tilde \Gamma_1(\varphi) - \tau_{\rm e}\}/T & \quad \mbox{for $0 \le \varphi < \tau_{\rm r}$}, \\
  \{\tilde \Gamma_2(\varphi) - \tau_{\rm e}\}/T & \quad \mbox{for $\tau_{\rm r} \le \varphi < \tau_{\rm e}$}, \\
  \{\tilde \Gamma_3(\varphi) - \tau_{\rm e}\}/T & \quad \mbox{for $\tau_{\rm e} \le \varphi \le T$},
 \end{array} \right. 
\end{equation}
where
\begin{align}
 \tilde \Gamma_1(\varphi) &= \frac{\tau_{\rm r}}{T}e^{\frac{\tau_{\rm e}}{\tau_{\rm r}}\varphi}
 + \left( -\varphi+\tau_{\rm e}-\frac{\tau_{\rm r}}{T}\right) e^{-\varphi}.\\
 \tilde \Gamma_2(\varphi) &=  \left( \varphi - \tau_{\rm r}+\frac{\tau_{\rm
 r}}{T}\right) e^{T-\varphi}
 + \left( -\varphi +\tau_{\rm e}- \frac{\tau_{\rm r}}{T}\right) e^{-\varphi},\\
  \tilde \Gamma_3(\varphi) & = \left( \varphi-\tau_{\rm r}+\frac{\tau_{\rm
 r}}{T}\right) e^{T-\varphi} 
  - \frac{\tau_{\rm r}}{T}e^{-\frac{\tau_{\rm e}}{\tau_{\rm r}}(T-\varphi)},
\end{align}
Alternatively, by introducing the phase $\phi$ ($0 \le \phi < 2 \pi$) as
\begin{equation}
 \phi = \omega \varphi,
\end{equation}
where $\omega=\frac{2\pi}{T}$, we obtain
\begin{equation}
 \dot \phi = \omega + K \Gamma(\phi-\phi'),
  \label{pm2}
\end{equation}
where
\begin{equation}
 \Gamma(\phi) = \omega \tilde \Gamma\left(\frac{\phi}{\omega}\right).
\end{equation}
For both Eqs.~\eqref{pm1} and \eqref{pm2}, $|K|$ values only determine the
time scale, thus we may set $K=1$ or $K=-1$ for positive and negative coupling, respectively, without loss of generality.
In Fig.~\ref{fig:aif_pm}, we show typical $Z(\varphi)$ and $\Gamma(\phi)$ functions.
There is a notable similarity between $\Gamma(\phi)$ functions obtained
experimentally and theoretically (Figs.~\ref{fig:exp_pm}(a) and
\ref{fig:aif_pm}(c)). In particular,
there is a rapid growth in a region of small $\phi$, and the slope is
steeper for a more relaxation oscillator, i.e., higher $U$ and
$\tau_{\rm e}$ values.

\subsection{Analysis}
\begin{figure}
 \includegraphics[width=8cm]{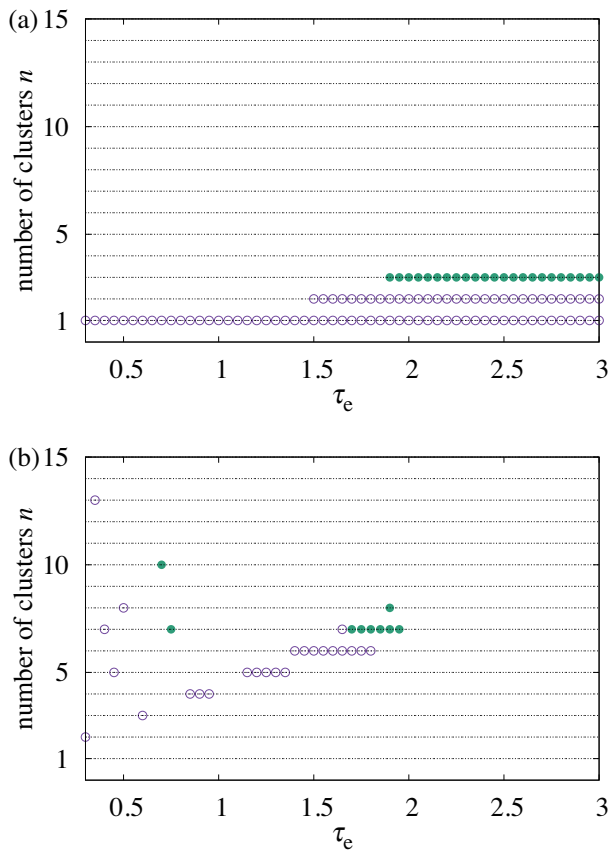}
 \caption{Stability diagram for the AIF system with global coupling. (a)
 positive coupling. (b) negative coupling. Each circle
 indicates that the $n$-cluster state is stable at the $\tau_{\rm e}$ value.
 Open and filled circles indicate that the largest nontrivial eigenvalue is
 a real and imaginary value, respectively.
 }
 \label{fig:stability}
\end{figure}

Now we analyze a system of globally coupled oscillators.
The system is given as
\begin{equation}
 \dot v_i = v_i + \frac{K}{N}\sum_{j=1}^N (v_j - v_i),
\end{equation}
and its corresponding phase model is obtained as
\begin{equation}
 \dot \phi_i = \omega + \frac{K}{N}\sum_{j=1}^N \Gamma(\phi_i-\phi_j),
  \label{pm_global}
\end{equation}
where $v_i$ and $\phi_i$ ($i=1,\ldots,N$) is the state and phase of
oscillator $i$, respectively.
As described in Appendix \ref{sec:stability}, the local stability of the balanced $n$-cluster states is determined by the nontrivial maximum
eigenvalue \cite{okuda93}. Figure \ref{fig:stability} shows a stability diagram.
Here, we only consider $n \le 10$ for simplicity.
With positive coupling ($K>0$),
at low values of $\tau_{\rm e}$ only the 1-cluster state is stable.
As the relaxation character increases $2, 3, 4$-cluster states
become progressively stable.
With negative coupling ($K<0$),
stable cluster states exist for small $\tau_{\rm e}$ values.
However, all cluster states are predicted to be unstable for large $\tau_{\rm e}$ values.
Numerical simulations with the AIF model indicate
that the average Kuramoto order parameter $\langle R \rangle$ is
vanishingly small for small $\tau_{\rm e}$ values.
There, in accordance with the stability analysis,
balanced cluster states were observed.
However, $\langle R \rangle$ begins to increase at
$\tau_{\rm e} \approx 2.0$ and
takes a large value (close to unity) as $\tau_{\rm e}$ increases.

\begin{figure}
 \includegraphics[width=8cm]{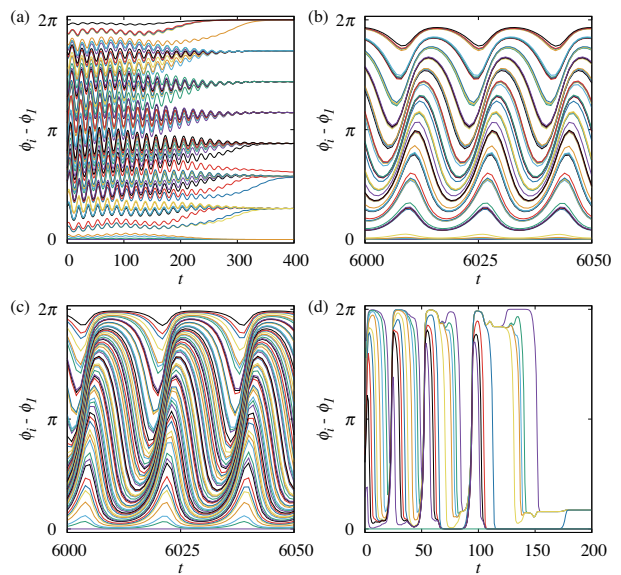}
 \caption{Time series of phase differences $\phi_i - \phi_1$ of the AIF system.
 (a) $\tau_{\rm e}=1.90$. (b) $\tau_{\rm e}=2.00$. (c) $\tau_{\rm
 e}=2.20$. (d) $\tau_{\rm e}=3.30$. $N=64$ except (d), where $N=10$ for
 better visibility. 
 }
 \label{fig:ts}
\end{figure}

To understand the emergence of synchrony, 
we first focus on its onset at $\tau_{\rm e}
\approx 2.0$. As shown in Fig.~\ref{fig:stability}(b), only the balanced
7-cluster state is stable just below $\tau_{\rm e} = 2.0$ and the state loses
stability at $\tau_{\rm e} \approx 2.0$. As also indicated in
Fig.~\ref{fig:stability}(b), the maximum nontrivial eigenvalue is
imaginary and its mode is associated with inter-cluster fluctuations
(see Appendix~\ref{sec:stability}). This implies that 
the balanced 7-cluster state loses its stability through a
Hopf bifurcation and the distribution of relative phases $\phi_i-\phi_1$
starts to oscillate after that. Figure \ref{fig:ts} shows the time
series of relative phases for different $\tau_{\rm e}$ values. 
At $\tau_{\rm e} = 1.8$ [Fig. \ref{fig:ts}(a)], the system converged to
the balanced 7-cluster state from a random initial condition, as
predicted. 
At $\tau_{\rm e} = 2.0$ [Fig. \ref{fig:ts}(b)], where no
balanced cluster states are predicted to be stable, the system converged to
a slightly scattered balanced cluster state, in which relative phases between
clusters oscillate with time. This state can be interpreted as a similar
state to those that bifurcate from the balanced $n$-cluster states via Hopf
bifurcations. 

\begin{figure}
 \includegraphics[width=7cm]{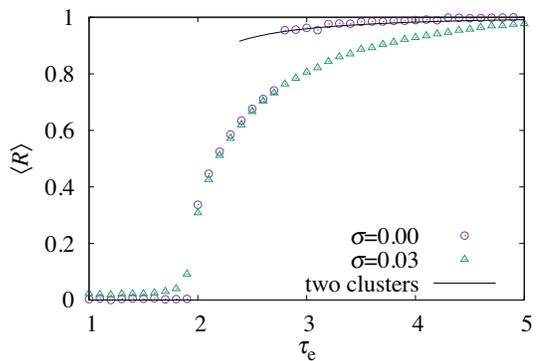}
 \caption{Average order parameter
 $\langle R \rangle$ versus $\tau_{\rm e}$ in the AIF system 
 ($N=64$) with negative coupling for noiseless $(\sigma=0.00)$ and noisy
 ($\sigma=0.03$) systems. The solid curve shows predicted $\langle R
 \rangle$ values, give by Eq.~\eqref{r_ss}.}
 \label{fig:kuramoto}
\end{figure}

For larger $\tau_{\rm e}$ values, no well-defined clusters are
observed. Instead, the oscillators form a noisy
cloud  similar to
Fig.~\ref{fig:exp_pm}(d)
in spite of the instability of the one-cluster state.
Figure \ref{fig:ts}(c) shows a typical time series of relative phases,
where the center of the cloud travels with time like a wave; i.e. each
oscillator enters and exits from a cloud repeatedly.
When $\tau_{\rm e}$ is further increased,
$\langle R \rangle$ suddenly jumps at $\tau_{\rm e} \approx 2.8$, as
shown in Fig.~\ref{fig:kuramoto}. Figure \ref{fig:ts}(d) shows typical
time series of relative phases for $\tau_{\rm e} > 2.8$.
The oscillators
split into two groups, and each group repeats aggregation and breakup.
Such a phenomenon is refereed
to as ``slow switching'', 
as the system slowly switches back and forth between
a pair of two
cluster states \cite{hansel93,kori01,kori03}.
This phenomenon occurs, because there are attracting heteroclinic cycles
between pairs of unstable out-of-phase two-cluster states. The
condition for the existence of attracting heteroclinic cycles
is obtained through the stability analysis of the two-cluster
states that are different from the balanced two-cluster states
(see Appendix \ref{sec:ss}). As a result, we find that the 
attracting heteroclinic cycles exist in our AIF system for large
$\tau_{\rm e}$ values. If the system converges to such a cycle, 
$\langle R \rangle$ is well approximated by
\begin{equation}
 \langle R \rangle \approx \sqrt\frac{1+\cos \Delta \phi}{2},
  \label{r_ss}
\end{equation}
where $\Delta \phi$ is the phase difference between two clusters.
$\Delta \phi$ is obtained as the solution to $\Gamma(\Delta \phi)=\Gamma(-\Delta \phi)$
(see Appendix \ref{sec:ss}), which is typically small, e.g., $\Delta \phi
\approx 0.3$ rad at $\tau_{\rm e}= 0.3$, thus high $\langle R \rangle$
values are predicted. In Fig.~\ref{fig:kuramoto}, predicted $\langle R
\rangle$ values are plotted as the solid curve, which is in a good
agreement with numerical $\langle R \rangle$ values for $\tau_{\rm e} >
2.8$. This result indicates that the heteroclinic cycles become attracting at
$\tau_{\rm e} \approx 2.8$. When heteroclinic cycles are nonattracting, 
it can be generally expected that attracting limit-cycles exist close to the
heteroclinic cycles. Actually, the phenomenon shown in Fig.~\ref{fig:ts}(d)
is rather similar to that in Fig.~\ref{fig:ts}(c), in particular before the
system gets very close to two cluster states (e.g., $t \approx
50$). Thus, our interpretation of noisy one-cluster state is a noisy
dynamics along heteroclinic cycles between 
unstable, saddle type cluster states. 

We also investigate the effect of noise. We consider
\begin{equation}
 \dot \phi_i = \omega + \frac{K}{N}\sum_{j=1}^N \Gamma(\phi_i-\phi_j)
  + \sigma \xi_i,
  \label{pm_global_noisy}
\end{equation}
where $\sigma$ is noise intensity and
$\xi_i(t)$ is white Gaussian noise with zero mean and unit variance.
The green triangles in Fig.~\ref{fig:kuramoto} show $\langle R \rangle$ values
obtained by numerical simulation of Eq.~\eqref{pm_global_noisy}. 
As seen, $\langle R \rangle$ values are similar to those in
the noiseless system for $\tau_{\rm e} < 2.8$. Effect of noise is
significant $\tau_{\rm e} \ge 2.8$ because noise inhibits the system
to get very close to unstable two-cluster states \cite{hansel93,kori01,kori03}.
Thus, with noise, the noisy one-cluster state persists even for
$\tau_{\rm e} \ge 2.8$. By a numerical simulation, we
confirmed qualitatively the same result is obtained for nonidentical
oscillators (i.e., $\omega$ in Eq.~\eqref{pm_global_noisy} is
$i$-dependent). These results indicate that
the noisy one-cluster state is robust against noise. 
Therefore, we interpret that the noisy one-cluster state observed in the
experiment is generated by an itinerant synchronization
involving unstable, saddle type cluster states. 

\section{Concluding Remarks}
In summary, we have shown that a noisy synchronized state can occur 
in a negatively coupled electrochemical oscillator system; this synchronized state was explained 
theoretically as an itinerant motion among unstable cluster states. This itinerant synchronization could further
contribute to the wide range of emergent collective behavior of physical, chemical,
and biological oscillators. 
We expect that similar phenomena can be reproduced in other oscillators
when they are close to homoclinic bifurcation. For example, the a Belousov-Zhabotinsky oscillatory reaction also exhibits phase response curve similar 
to the predictions of AIF model \cite{Chen:2016oq}. Neural models
are good candidates as well, because many of them, such as the Morris-Lecar model  \cite{izhikevich07}, exhibit homoclinic
bifurcation.

Many integrate-and-fire models proposed
previously assume instantaneous resetting, which yields discontinuous
$v(t)$ \cite{lewi09,ermentrout86,kuramoto91,kori01,Olmi:2010fj}.
In contrast, in the AIF model, we have 
introduced the resting process with finite
period $\tau_{\rm r}$, which enables us to consider continuous $v(t)$. This feature is not only
natural but also a great advantage in mathematical and numerical
treatments because then our model
has continuous flow at any time.
Thereby, delicate problems due to discontinuity can be avoided.
For example, it is important that the interaction function $\Gamma(\phi)$
has continuous derivative at $\phi=0$, i.e., $\Gamma_1'(0)=\Gamma_1'(2\pi)=-\tau_{\rm e}$ for nonvanishing $\tau_{\rm r}$, because
$\Gamma'(0)$ plays a vital role in determining many synchronous states.

\acknowledgments{H.K. acknowledges supports from
MEXT KAKENHI Grant Number 15H05876.
I.Z.K. acknowledges support from National Science
Foundation CHE-1465013 grant. 
One of the authors, H.K., met John Hudson at Fritz Haber Institute in
Berlin in 2003, and after that, had many occasions to collaborate with
him. All of them were great experiences for H.K., through which H.K
learned a lot about how to work on science and enjoy collaborations.}

\appendix
\section{Stability of the balanced cluster states} \label{sec:stability}
We briefly summarize the stability analysis of the balanced cluster
states \cite{okuda93}.
In the phase model given by Eq.~\eqref{pm_global},
the balanced $n$-cluster state always exist for any $\Gamma$.
In the balanced $n$-cluster state, $N/n$ oscillators make a point
cluster and these oscillators take the same phase 
$\Phi_k$ ($k=0,1,\ldots, n-1$), given by
\begin{equation}
 \Phi_k = \Omega t + \frac{2\pi k}n,
  \label{cluster_k}
\end{equation}
where $\Omega$ is the actual frequency. By substituting
Eq.~\eqref{cluster_k} into Eq.~\eqref{pm_global}, we obtain
\begin{equation}
 \Omega = \frac{1}{n} \sum_{k=0}^{n-1} \Gamma\left(\frac{2\pi k}{n}\right).
\end{equation}
By solving the eigenvalue problem for the corresponding stability
matrix, we obtain $N$ eigenvalues as
\begin{align}
 \tilde \lambda &=\frac{1}{n} \sum_{k=0}^{n-1} \Gamma'\left(\frac{2\pi
 k}{n}\right), \\
 \lambda_p &=\frac{1}{n} \sum_{k=0}^{n-1} \Gamma'\left(\frac{2\pi
 k}{n}\right) \left( 1 - e^{-{\rm i} 2\pi k p/n}\right),
\end{align}
where 
the former has $N-n$ multiplicity and is associated with
intra-cluster fluctuation; and the latter has $1$ multiplicity
for each $p$ ($p=0,1,\ldots,n-1$) and is associated with inter-cluster
fluctuation. There is one trivial eigenvalue $\lambda_0=0$, which is
associated with uniform phase shift. 
The balanced $n$-cluster state is linearly stable if and only if
all the remaining eigenvalues have negative real parts.

The derivative of $\tilde \Gamma(\varphi)$ is 
\begin{equation}
 \tilde \Gamma'(\varphi) = \left\{
 \begin{array}{cl} 
  \tilde \Gamma'_1(\varphi) /T & \quad \mbox{for $0 \le \varphi \le \tau_{\rm r}$}, \\
  \tilde \Gamma'_2(\varphi) /T & \quad \mbox{for $\tau_{\rm r} \le \varphi \le \tau_{\rm e}$}, \\
  \tilde \Gamma'_3(\varphi) /T & \quad \mbox{for $\tau_{\rm e} \le \varphi \le T$},
 \end{array} \right. 
\end{equation}
where
\begin{align}
 \tilde \Gamma'_1(\varphi) &=  \frac{\tau_{\rm e}}{T}e^{\frac{\tau_{\rm e}}{\tau_{\rm r}}\varphi}
 - \left( -\varphi + \tau_{\rm e}-\frac{\tau_{\rm r}}{T}+1 \right) e^{-\varphi}.\\
 \tilde \Gamma'_2(\varphi) &=  -\left( \varphi - \tau_{\rm r}+\frac{\tau_{\rm
 r}}{T}-1\right) e^{T-\varphi} \nonumber \\
  &- \left( -\varphi + \tau_{\rm e}-\frac{\tau_{\rm r}}{T}+1\right) e^{-\varphi},\\
 \tilde \Gamma'_3(\varphi) &= -\left( \varphi - \tau_{\rm r}+\frac{\tau_{\rm
 r}}{T}-1 \right) e^{T-\varphi} 
  - \frac{\tau_{\rm e}}{T}e^{-\frac{\tau_{\rm e}}{\tau_{\rm r}}(T-\varphi)}.
\end{align}
Note $\Gamma'(\phi) = \frac{d \varphi}{d \phi}\frac{d}{d\varphi}\left(\omega
\tilde \Gamma(\varphi)\right) = \tilde \Gamma'\left(\frac{\phi}{\omega}\right)$.

\section{Existence and stability of two-cluster states} \label{sec:ss}
We briefly summarize the existence and stability analysis of the
two cluster states and the condition for the existence of attracting
heteroclinic cycles between a pair of two cluster states
\cite{hansel93,kori01}.
There is a family of two-cluster states in Eq.~\eqref{pm_global}, in
which $q N_1$ oscillators and $(1-q)N$ oscillators form point clusters.
Let $\phi_{\rm A}$ and $\phi_{\rm B}$ be the phase of these clusters.
The phase difference $\Delta \phi = \phi_{\rm A} - \phi_{\rm B}$ is
obtained as the solution to
\begin{equation}
 (2q-1) \Gamma(0) + (1-q)\Gamma(\Delta \phi) - q\Gamma(-\Delta \phi) =
  0.
  \label{Delta_phi}
\end{equation}
The eigenvalues of the corresponding stability matrix are
\begin{align}
 \lambda_1 &= K{q \Gamma'(0) + (1-q) \Gamma'(\Delta \phi)},\\
 \lambda_2 &= K{(1-q) \Gamma'(0) + q \Gamma'(-\Delta \phi)},\\
 \lambda_3 &= K{(1-q) \Gamma'(0) + q \Gamma'(\Delta \phi)},
 \label{lambda_ss}
\end{align}
with multiplicities $Nq-1, N(1-q)-1, 1$, respectively.
There is also one trivial eigenvalue $\lambda_0=0$.
Eigenvalues $\lambda_1$ and $\lambda_2$ are associated with
intra-cluster fluctuation, and $\lambda_3$ is associated with
inter-cluster fluctuation.
For generic $\Gamma$, many of the two-cluster states are saddles;
i.e., only a part of eigenvalues are negative. 
Nevertheless, such saddle states are meaningful because pairs of the
two-cluster states form attracting heteroclinic cycles and the system
may approach one of them. From here, for simplicity, we 
only consider two-cluster states with $q=\frac{1}{2}$.
There are a pair of two-cluster states with the phase differences $\pm \Delta
\phi$. The heteroclinic cycle can be formed
between this pair of cluster states, if
\begin{equation}
 \lambda_1>0, \lambda_2<0, \lambda_3<0.
  \label{stability1}
\end{equation}
Furthermore, the cycle can be attracting, if
\begin{equation}
 \left| \frac{\lambda_1}{\lambda_2} \right| \lesssim 1.
  \label{stability2}
\end{equation}
The solid line in Fig.~\ref{fig:kuramoto} is plotted in the following
manner. For given $\Gamma$, Eq.~\eqref{Delta_phi} is solved numerically
to find $\Delta \phi$. Using this $\Delta \phi$ value, we check
Eqs.~\eqref{stability1} and \eqref{stability2}. If both stability
conditions are satisfied, we plot a $R$ value given by Eq.~\eqref{r_ss} in
Fig.~\ref{fig:kuramoto}. 


\end{document}